\documentclass{iauc}
\usepackage{graphicx}
\def\gsim{\;\rlap{\lower 2.5pt
 \hbox{$\sim$}}\raise 1.5pt\hbox{$>$}\;}
\def\lsim{\;\rlap{\lower 2.5pt
   \hbox{$\sim$}}\raise 1.5pt\hbox{$<$}\;}
\pubyear{2005}
\volume{199}
\pagerange{1--}
\jname{Probing Galaxies through Quasar Absorption Lines}
\editors{P. R. Williams, C. Shu, and B. M\'{e}nard, eds.}

\title[Intergalactic enrichment: theory vs. observations] 
{Observational Tests of Intergalactic Enrichment Models}

\author[Aguirre \& Schaye]   
{Anthony Aguirre$^1$ \and
Joop Schaye$^2$}

\affiliation{$^1$Department of Physics, UCSC, Santa Cruz CA 95064 \break 
email: aguirre@scipp.ucsc.edu\\[\affilskip]

$^2$Leiden Observatory, P.O. Box 9513, 2300 RA Leiden, The Netherlands
\break 
email: schaye@strw.leidenuniv.nl }

\begin{document}

\maketitle

\begin{abstract}

We summarize recent results assessing the carbon and silicon
abundances of the intergalactic medium (IGM) using the `pixel optical
depth' technique. We briefly discuss the implications of these results
for models of intergalactic enrichment, focusing on distinguishing
`early' $z \gg 4$ enrichment by the first generations of stars and
objects from `late' enrichment by $2 \lsim z\lsim 5$ Ly-break
galaxies.  We then discuss the comparison of observed QSO spectra to
simulated spectra generated from cosmological simulations that
self-consistently include enrichment, and draw qualitative
implications for the general picture of intergalactic enrichment at $z
\gsim 2$. \\ 
Intergalactic medium -- Galaxies: formation

\end{abstract}

\firstsection 
\section{Introduction}

The widespread existence of metals outside of galaxies has been known
for a decade (e.g., Cowie et al.\ 1995) by their absorption lines in
high-$z$ QSO absorption spectra, but their origin remains somewhat of a
mystery.  Were they created in-situ, or removed from galaxies by
winds, galaxy dynamics, dust flows, or some other mechanism? And {\em
when}?

This question has been very challenging in part because of the paucity
of {\em detailed} information on the distribution of intergalactic
metals. Fortunately over the past few years several groups have been
employing large samples of Keck and VLT high-resolution spectra to
infer a more detailed view of the observed enrichment (Schaye et
al. 2003, hereafter S03; Adelberger et al.\ 2003;
Pichon et al.\ 2003; Pieri \& Haehnelt 2004;
Aguirre et al.\ 2004, hereafter A04; Simcoe et al.\ 2004; Aracil et
al. 2004).  Meanwhile, cosmological simulators have begun to
incorporate metal generation and transport in their simulations and
make comparisons to the observed absorption spectra (Theuns et
al. 2002; Cen et al.\ 2004; Aguirre et al.\ 2005, hereafter A05).
Herein we will review the findings of our group regarding
intergalactic abundances, and discuss in a preliminary way what those
findings suggest about the mechanism of enrichment. In this exercise,
it is useful to have in mind two basic paradigms.

The first we might call ``late enrichment'', wherein the relatively
massive Ly-break galaxies observed at $z \lsim 6$ are
responsible. This model has the virtue that strong winds are in fact
observed in such galaxies (e.g., Pettini et al.\ 1998), and that much
more star formation takes place at $z \lsim 6$ than at $z \gsim
6$. Its deficiencies are that the regions near such galaxies comprise
a small fraction of the cosmic volume -- so spreading metals widely
enough may be difficult -- and that the high outflow velocities
necessary to escape the galaxies' potential wells will lead to high
post-shock gas temperatures in the IGM (as well as structural changes
in the IGM near galaxies) that may be incompatible with observations
of the Ly$\alpha$ forest.

In a second scenario, ``early enrichment'', the very earliest
generation of protogalaxies and stars (including Population III) at $z
\gg 5$ pollute the IGM.  As these are more numerous than the Ly-break
galaxies they might affect a larger filling factor, and they would
provide the IGM with more time to recover from the effect of
winds and appear relatively undisturbed (e.g., Madau, Ferrara \& Rees
2001) by $z\sim 3$. For present purposes we will model such enrichment
as metals 
``sprinkled'' into the IGM (without affecting the gas at all) at $z >
5$, with no subsequent metal evolution.

How might we distinguish these scenarios (or others) using
observations of the IGM at $z \lsim 4$?  There are a number of ways.
1) If the observed enrichment in a given environment {\em evolves}
then this directly indicates that enrichment by $z < 4$ galaxies is
important.  2) The spatial distribution of metals in the two scenarios
should differ -- though in a way that requires careful modeling.  3)
In late enrichment many metals may lie in warm/hot collisionally
ionized gas rather than in warm ($\sim 10^4\,$K) photoionized gas, so
probing the temperature of the observed metals would be very useful.
4) Abundance ratios of the IG metals would aid in determining the
enriching sources.  5) The total amount of intergalactic metal may be
compared to that expected in different enrichment models.

Fortunately, large samples of data as well as advances in the way the
data is analyzed now makes this possible.  In particular, the ``pixel
optical depth'' method, pioneered by Cowie \& Songaila (1998), and
tested, extended, and employed by Aguirre et al.\ (2002, hereafter
A02), S03, A04 and A05 allows significant progress in all of these
directions.

There are two basic means of proceeding in this aim; inferring
abundances from the observations as directly as possible and comparing
to these models, or generating models and comparing these directly to the
observations.  We have taken both approaches, as described in Sections
2 and 3, respectively.

\section{The inferred intergalactic abundances}

The pixel technique is described in detail in A02, S03, and A04. At
the most basic level, one plots the optical depths for pixels in a
wavelength region where some transition is expected, against those in
wavelengths corresponding to absorption by the same gas in another
transition. Correlation implies detection: for example, correlation
between CIV optical depth $\tau_{\rm CIV}$ and Ly$\alpha$ optical
depth $\tau_{\rm Ly\alpha}$ indicates the presence of carbon.  The
results -- and implications for the enrichment scenarios -- of
applying a suite of techniques based on this idea to a sample of 19
Keck and UVES spectra is summarized below.

{\bf\em Metallicity is inhomogenous and density-dependent.}  By
assuming an ultraviolet background (UVB) model (our fiducial choice is
that of Haardt \& Madau 2001 including both quasars and galaxies), and
that the relation between $\tau_{Ly\alpha}$ and gas overdensity
$\delta$ is as given by hydrodynamics simulations, we can covert
$\tau_{\rm CIV}/\tau_{\rm Ly\alpha}$ vs. $\tau_{\rm Ly\alpha}$ into
[C/H] versus $\delta$ (see A02).  The result (S03) is that the median
[C/H] depends on density:
$$
[{\rm C}/{\rm H}] = -3.47_{-0.06}^{+0.07}+ 0.65_{-0.14}^{+0.10}\times(\log\delta-0.5),
$$ where here and elsewhere errors are $1\sigma$.  Whether enrichment
is early or late, it is unsurprising that high-density regions are
more enriched, but this trend had not previously been established.
This result (as well as some others below) does depend on the UVB
model chosen; for example, a harder UVB (e.g., as from quasars only)
would imply a higher metallicity overall, and less of a trend of metallicity
with density.  However, UVBs very different from our fiducial
model are strongly disfavored.  For example, if the UVB is much harder,
then the inferred Si/C ratio (see below) is implausibly high, and the
{\em mean} metallicity of the IGM actually increases with decreasing
density, which seems unlikely; or the UVB is much softer, the
inferred O/C or O/Si will be implausibly high.

Beyond the medians, the {\em full} distribution of $\tau_{\rm
CIV}/\tau_{\rm Ly\alpha}$ for a given $\tau_{\rm Ly\alpha}$ can be
used to determine the scatter in [C/H] at a given density (S03).  This
reveals that [C/H] is well-fit by a normal distribution with a width
of $\approx 0.8$\,dex. This rules out any truly uniform enrichment
scenario, and can provide strong constraints when compared in detail
to enrichment models (see below).

{\bf\em Metallicity does not appear to evolve significantly.}  Perhaps
even more interestingly, [C/H] at a given gas density appears
independent of redshift (S03): a simultaneous fit of the $\delta-$ and
$z-$ dependence for all data yields $[{\rm C}/{\rm H}] \propto
(0.08\pm{0.10})\times(z-3),$ and for any given cut in density our
results are compatible with no evolution.\footnote{Note that even if
all enrichment were completed by $z=4$, metallicity would still evolve
because overdense regions increase their overdensity with time, and
this would cause a flattening of the [C/H] vs $\delta$ trend.  This
effect is compatible with our measurements and is in fact suggested by
our analysis, but only at $\approx 1\sigma$ significance (see S03,
Sec. 8.1)} Clearly this suggests early enrichment and indicates that
the metals we see were in place by $z \sim 4$. Metals could in
principle be added to the IGM after $z\sim 4$, but only in such a way
that they cannot be observed in CIV absorption.

{\bf\em The observed metals are largely in warm, photoionized gas.}
The CIII and SiIII lines fall in the Ly$\alpha$ forest and, lacking
doublets, are difficult to identify using line-fitting. They can,
however, be measured using the pixel technique -- the forest appears
as a contaminant that can be reliably subtracted.  Because the ratios
CIII/CIV and SiIII/SiIV are quite sensitive to both the density and
temperature of the absorbing gas (but rather insensitive to the
spectral shape of the UVB), we can use them to strongly constrain the
state of the gas giving rise to the observed absorption. Our results
(S03; A04) indicate that the relatively high-density CIV- and SiIV
absorbing gas is at $T < 10^{4.9}\,K$.  Moreover, the observed
$\tau_{\rm CIII}/\tau_{\rm CIV}$ and $\tau_{\rm SiIII}/\tau_{\rm
SiIV}$ ratios are fully compatible with those predicted from numerical
simulations with {\em no} feedback in which metals are added to the
IGM in accord with the carbon distribution inferred from the
observations (S03; A04; A05).  As for the metallicity vs. redshift,
this indicates that our results are compatible with the early
enrichment ``sprinkled metals'' model; if carbon or silicon is added
to the IGM by later generations of galaxies, then it must not absorb
strongly in CIV (i.e., it must be too hot and/or too low-density,
and/or too low-metallicity).

{\bf\em Abundance ratios are consistent with no evolution.} We can use
ions of different elements to infer relative abundances: for example,
from $\tau_{\rm SiIV}/\tau_{\rm CIV}$ we can infer [Si/C] and from
$\tau_{\rm OVI}/\tau_{\rm CIV}$ we can infer [O/C]. In addition, the
distribution of $\tau_{\rm SiIV}/\tau_{\rm CIV}$ at a given $\tau_{\rm
CIV}$ constrains the {\em scatter} in [Si/C]. If a Haardt \& Madau
(2001) UVB is assumed, the analysis yields [Si/C]$=0.77\pm 0.05$.  (As
for the carbon abundance, this result depends on the UVB: a Haardt \&
Madau UVB with no galaxies would yield [Si/C] $\simeq 1.5$, and softer
UVBs would lead to [Si/C] $< 0.75$.) There is no evidence for
evolution, density-dependence, or scatter in [Si/C].  The [Si/C] value
we obtain is probably within systematic errors\footnote{Specifically,
there are uncertainties in the recombination rates of SiIV (A04) and
in the importance of local sources of UV photons (Schaye 2004) that
could affect the result at this level.}  of the [Si/C]$\sim 0.5$
expected from Type II supernovae; alternatively, the high value {\em
might} suggest some contribution from massive, metal-free Pop. III
stars which can (according to theoretical models), give as high as
[Si/C]$ \sim 1.3$ (see the compilation by Qian \& Wasserburg
2005). Note that making the UVB softer would be an easy way to
reconcile our [Si/C] with type II yields, but that preliminary
analysis indicates supersolar [O/C] for the same UVB, and [O/C] would
{\em increase} with a softer UVB; so it seems that either O and/or Si
is highly overabundant with respect to C.

What does this say about the enrichment scenario? Type II yields would
be unsurprising from either early or late enrichment (as both would
probably lack significant contributions from Type Ia or AGB
enrichment) while Pop. III yields would be expected only to contribute
significantly in the early scenario. The lack of evolution in [Si/C]
is more evidence against anything evolving significantly, as
predicted by the early scenario.  Finally, the lack of scatter in
[Si/C] suggests that the observed carbon and silicon result from the
same nucleosynthetic sources, and are distributed by the same
processes.

{\bf\em There is a lot of metal in the IGM.}  By combining the center
and width of the metallicity distribution at a given density with the
fraction (as predicted by simulations) of the IGM at each density, we
can add up the contribution of the observed metals to the total metal
content of the IGM.  Performing this calculation for our observations
over the density range $-0.5 \le \log\delta \le 2.0$ yields $
\Omega_{\rm C}=2.3\times 10^{-7} $ for our fiducial UVB.  Using also a
uniform value of [Si/C]=0.77 gives $ \Omega_{\rm Si}=3.4\times
10^{-7}.  $

It is interesting to compare this to a rough estimate of the amount of
silicon expected to be produced by stellar nucleosynthesis at $z \gsim
3$:
$$
\Omega_{\rm Si,exp}=10^{-6} \left({Y\over 0.02}\right) \left({f_{\rm Si}\over 0.05}\right) \left({\Omega_*\over 0.001}\right),
$$ where $Y$ is the metal mass/stellar mass formed, $f_{\rm Si}$ the
the metal mass fraction in silicon, and $\Omega_*$ is the stellar
density at $z\sim 3$.  This indicates that a substantial fraction of
all metals expected to be produced by $z \sim 3$ reside in the IGM.
Note that this may cause difficulties if all of the enrichment
happened at $z \gg 4$, as the total stellar mass formed at such early
times may be insufficient to produce the requisite metalllicity unless
the early star formation rate is assumed to be extremely high (e.g.,
Qian \& Wasserburg 2005).

\section{Comparing observed to simulated spectra}

The abundances inferred from the CIV and SiIV optical depths are
clearly compatible the early model in which metals are
sprinked into the gas at $z \gsim 4$ without affecting the
gas. However, the the conversion of optical depths into C and H
densities makes use of simulations {\em without feedback}; it is
therefore self-consistent only in the early scenario.

It is thus useful to employ a complementary approach of generating
spectra from simulations with enrichment -- including the effect of
the outflows on the IGM -- and comparing these to the observed spectra
using the pixel technique.  In a first investigation of this sort
(A05) we have performed this analysis using two sets of
high-resolution particle-pased hydrodynamic simulations (Theuns et
al.\ 2002; Springel \& Hernquist 2003) with different feedback
prescriptions that drive strong winds at $z \gsim 2$ as per the late
enrichment scenario.  This led to several general and qualitative
conclusions:

{\bf\em Simulation with feedback under-estimate CIV and CIII
absorption.} As mentioned above, the observed CIV/HI and CIII/CIV
ratios are well-reproduced by non-feedback simulations if we impose a
metallicity distribution on the gas as inferred from the
observations. For CIV/HI, this is just a consistency check, but the
CIII/CIV comparison is a genuine success.  However, when spectra are
generated from self-consistent enrichment simulations, they produce
far too small values of $\tau_{\rm CIV}/\tau_{\rm HI}$ and $\tau_{\rm
CIII}/\tau_{\rm CIV}$, for all of our UVB models (though softer UVBs
make the discrepancy smaller).

{\bf\em Metal rich gas in the simulations is too hot and too
low-density.} The problem arises because essentially all of the
intergalactic and enriched gas in both simulations exists in 
low-density, high-temperature ($10^5\,K \lsim T \lsim 10^6\,K$)
bubbles.  Since both the CIV/C ratio and the CIII/CIV ratio fall off
quickly with both increasing temperature $T \gsim 10^5\,$K, and
decreasing density $\delta \lsim 10$ the gas becomes nearly invisible
in CIV, and where CIV is detected there is virtually no accompanying
CIII.

{\bf\em Metal line cooling should be important.} There is a
potentially important effect that could ameliorate the aforementioned
problem, which is that cooling by metals was not included in the
simulations, but could be important due to the high metallicity of the
gas in the bubbles.  Calculating the cooling time using the Sutherland
\& Dopita (1996) tables indicates that much of the metal-rich gas
could plausibly cool. Correctly treating metal cooling in cosmological
simulations will be quite challenging because it depends on the very
local metallicity (and thus crucially on how well metals mix) and
equilibrium cooling may not be a good approximation.  But it can be
crudely modeled by assuming that gas cools to $T\approx 10^4\,$K if
its (Sutherland \& Dopita) cooling time is shorter than the Hubble
time. In this case much more CIV absorption is present and the {\em
median} $\tau_{\rm CIV}/\tau_{\rm HI}$ observed could plausibly be
matched by the simulations.

{\bf\em Nonetheless metals appear too low-density and too inhomogeneous.}
While including metal cooling might lead to a sufficient increase in
CIV absorption, it does {\em not} appear to fix the difficulty in
reproducing CIII/CIV.  This can be traced to the fact that the metal-rich gas has
too low density to produce the observed ratios.  This problem may also
be lessened by a proper treatment of metal cooling, as the
cooling would also affect the gas dynamics, allowing gas to contract
as it cools. But there is yet another problem, which is that although
the {\em median} $\tau_{\rm CIV}/\tau_{\rm HI}$ can be roughly
reproduced by the simulations, the spread in $\tau_{\rm CIV}$ at a
given $\tau_{\rm HI}$ cannot.  This can be traced to the fact that
metals in the simulations are {\em too inhomogeneous}, which is turn
results from a relatively small fraction of gas particles being
enriched.  This problem seems unlikely to improve from a correct
treatment of metal cooling.  
\section{Conclusions}

What are we to make of the question of early vs. late enrichment?  Our
analysis presents a bit of a conundrum.  On the one hand, all of our
observations of intergalactic enrichment appear perfectly consistent
with metals being in place in the IGM at $z \gsim 4$ without
disturbing the IGM (with the possible exception of the overall metal
mass, which may be challenging for $z >> 4$ enrichment to reproduce).
Moreover, simulations of enrichment caused by strong feedback fail to
reproduce the observations in a number of ways. On the other hand,
essentially all $z \gsim 2$ galaxies appear to be driving winds that
seem likely to escape into the IGM -- where do their metals go?
Furthermore, groups and clusters at $z \sim 0$ exhibit a metallicity
of $Z\sim {1\over 3}Z_\odot$ that is much higher than that of the IGM
at $z \sim 3$, some Ly$\alpha$ clouds have $Z\sim 0.1Z_\odot$ at $z
\lsim 0.5$ (Prochaska et al.\ 2004), and reasonable estimates of the
mean $z=0$ metallicity are $\sim 5-10\times$ higher than that inferred
at $z\sim 3$ (Finoguenov et al. 2003\footnote{They calculate
$\Omega_{Z,IGM}=3.3-7\times10^{-5}$ in clusters, OVI systems and the
Ly$\alpha$ forest, corresponding to $Z\approx 0.05-0.1 Z_\odot$ in the
IGM.}).  Thus it seems that either metallicity evolution occurs
between $z\sim 3$ and $z\sim0$, or we are failing to detect some
metals at $z\sim3$, or both.

Can this conflict be reconciled?  Perhaps.  It seems plausible that
galaxies forming at {\em all} redshifts $z \gsim 2$ contribute to
intergalactic metallicity: the metals we {\em see} in the Ly$\alpha$
forest via quasar spectroscopy were almost entirely in place by $z
\sim 4$ and result from a rather early enrichment phase that left the
IGM relatively undisturbed, while metals added at $2 \lsim z \lsim 4$
are largely hidden from view in hot, low-density gas with a relatively
small filling factor that preferentially avoids the dense filaments.
This latter enrichment could, through gravitational infall, later
appear as a portion of the high metal mass gas in clusters and groups
(the rest being provided by processes occurring after the
groups/clusters can be said to have formed). Whether such a scenario
can be made to work in detail is an open question.

While a lot of progress has been made in understanding the mechanism
of intergalactic enrichment, there are a number of important avenues
to pursue toward assembling a comprehensive picture of the process:
\begin{itemize}
\item Where possible, more ions (e.g., OVI and NV) should be analyzed
to provide stronger constraints on the UVB model, on relative
abundances, and on the possible importance of collisionally
ionization. Such work is currently underway.
\item Simulations of `late' enrichment that include metal cooling must
be evaluated, with particular attention to how well metal cooling is
captured, and what effect it has on the observed ionic ratios. It
would be advantageous to compare grid-based with particle-based
simulations of enrichment as well. Several groups will undoubtedly be
performing such analyses over the next several years.
\item Modeling and simulations of individual wind-driving galaxies
embedded in a realistic IGM are vital for understanding how winds will
really propagate into the IGM, and the link between enrichment and
galaxy formation processes.
\item More work on `early' enrichment is also important: is it really
possible to enrich the IGM without any real effect on the gas?  What role could
population III play?
\item Enrichment mechanisms differing from the canonical
supernova-driven wins are worth investigating.  For example, what
about outflows of dust (Aguirre et al.\ 2001; Bianchi \& Ferrara 2005)
or driven by dust (Murray et al.\ 2005)? And how much of the
group/cluster metals could be explained by dynamical processes
rather than outflows?

\end{itemize}

\label{sec:concl}

\end{document}